\newcommand{\sigmaq}{\Sigma q}
\newcommand{\sigman}{\Sigma n}

\newcommand{\qi}{q_{i}}  

\newcommand{\hi}{\phi_i}
\newcommand{\hj}{\phi_j}

\newcommand{\etai}{\eta_1}
\newcommand{\etaj}{\eta_{2}}

\newcommand{\lj}{l_j}
\newcommand{\lk}{l_k}

\newcommand{\nj}{n_{j}}
\newcommand{\nk}{n_{k}}

\newcommand{\qj}{q_{j}}
\newcommand{\qk}{q_{k}}

\newcommand{\er}{e_r}
\newcommand{\lr}{l_r}
\newcommand{\dr}{d_r}
\newcommand{\ur}{u_r}
\newcommand{\qr}{q_{r}}  
\newcommand{\nnr}{n_{r}}
\newcommand{\hr}{\phi_r}

\documentclass[11pt,twoside,dvips]{article}
\usepackage{verbatim}
\usepackage{amssymb}
\usepackage{amsfonts}
\usepackage{amsmath}
\usepackage{pifont}
\usepackage[spanish,english]{babel}
\usepackage[utf8]{inputenc}
\usepackage[pdftex]{graphicx}
\usepackage{verbatim}
\usepackage{epsfig}
\usepackage{bm}
\usepackage{longtable}
\usepackage{fancyhdr}
\usepackage{csquotes}
\usepackage{placeins}
\begin{document}

\fancypagestyle{plain}{%
\fancyhf{}%
\fancyhead[LO, RE]{XXXVIII International Symposium on Physics in Collision, \\ Bogot\'a, Colombia, 11-15 September 2018}}

\fancyhead{}%
\fancyhead[LO, RE]{XXXVIII International Symposium on Physics in Collision, \\ Bogot\'a, Colombia, 11-15 September 2018}

\title{Non-universal  electroweak extensions of the standard model and the scotogenic models.}
\author{Dario J. Blandon$\thanks{%
e-mail: dario.blandon@udea.edu.co}$ $^{a}$, Diego A. Restrepo$\thanks{%
e-mail: restrepo@udea.edu.co}$ $^{a}$, William A. Ponce$\thanks{%
e-mail: willian.ponce@udea.edu.co}$ $^{a}$,
Eduardo Rojas$\thanks{%
e-mail: eduro4000@gmail.com}$ $^{b}$, 
  \\ $^{a}$Instituto de Fisica, Universidad de Antioquia, \\ Medellin,Antioquia
\\ $^{b}$Departamento de F\'isica, Universidad de 
Nari\~no, \\  A.A. 1175, San Juan de Pasto, Colombia}
%EndAName
\date{}
\maketitle

\begin{abstract}
In order to analyze some low energy experimental anomalies, 
we charge with a non-universal   $U(1)'$ gauge symmetry
the standard model fermions, taking as a starting point
the well-known scotogenic model. 
In order to have  non-trivial solutions to the anomalies 
and the Yukawa constraints,  we add 
three neutral singlet Dirac fermions.  
We have found two possible non-universal solutions which, 
as a matter of principle, are suitable to analyze 
family-dependent experimental anomalies.
\end{abstract}

\section{Introduction}
In the SM the couplings of the photon and the  $Z$ boson to the standard model~(SM) fermions are universal, 
that is to say, the couplings of these bosons to the corresponding fermions
in each family are the same, having as a consequence that couplings 
of these bosons to the fermions of the SM continue to be diagonal after the
mixing between fermions~\cite{Langacker:2000ju}.
However, there are several  observables at low energies for which their 
experimental values get apart from those  predicted by the SM. 
In many of these cases the best models to fit these anomalies are non-universal 
electroweak extensions to the SM;  being this feature the main motivation for the present work.   
Some of the observables that show deviations 
from the SM are: 
the anomalous magnetic moment of the muon~\cite{Bennett:2006fi}, 
the beryllium anomaly~\cite{Krasznahorkay:2015iga}, 
MiniBooNE~\cite{Aguilar-Arevalo:2018gpe}, 
Gallium solar neutrino experiments
 GALLEX
\cite{Anselmann:1994ar,Hampel:1997fc,1001.2731} and
 SAGE~\cite{Abdurashitov:1996dp,hep-ph/9803418,nucl-ex/0512041,0901.2200},
NuTeV~\cite{Zeller:2001hh}, LSND~\cite{Aguilar:2001ty},
and the Reactor anomaly~\cite{Mueller:2011nm}.  
There is also an increasing interest in a number of anomalies in semileptonic $B$ decays  reported 
by the  LHCb collaboration and other 
experiments~\cite{Aaij:2014pli,Aaij:2014ora,Aaij:2013qta,Aaij:2015oid,Wehle:2016yoi,Aaij:2013aln,Aaij:2015esa}. 

On the other hand, in neutrino physics  the scotogenic models are compelling  from a  
theoretical point of view, because they manage to link the dark matter
and and the mechanism to generate the neutrino masses~\cite{Ma:2006km}.
Following well-established methods to induce
residual symmetries from $U(1)$-invariant theories~\cite{Krauss:1988zc,Petersen:2009ip},
within the framework of higher
gauge symmetries, the dark matter stability can be explained 
through residual symmetries at low energies~\cite{Sierra:2014kua,Walker:2009en,Batell:2010bp,Lindner:2013awa}.
In these models   the neutrino masses  are generated radiatively via the effective the Weinberg operator, 
in such a way  that the mechanism of generation of masses of neutrinos
and the dark matter are related to each other. 
As it has been proposed in some recent works, the stability 
of dark matter can be explained by residual symmetries of gauge groups; 
avoiding the need to impose  ad hoc symmetries. 
This procedure has been implemented in several models and in particular 
in some scotogenic models. 

These results motivate the study of non-universal models
and in particular those models 
that adjust some of these anomalies simultaneously. 
Motivated by this phenomenology, and bearing in mind that many of these anomalies suggest non-universal models,
 we extend the electroweak 
 sector of the SM with an additional symmetry $U(1)$ with non-universal charges to 
 the fermions of the SM.   
Additionally, we want our model to be able 
to explain masses of neutrinos and dark matter stability.
As we mentioned earlier the scotogenic paradigm is the preferred scenario in these cases. 

 The manuscript is organized as follows: in section~\ref{sec:anomalies} write the anomaly equations and  
the restrictions coming from the Yukawa terms and the 1-loop neutrino self-energy diagram. 
In section~\ref{sec:results} we will write our results.

\section{Anomaly and Yukawa constraints}
\label{sec:anomalies}
Following ref~\cite{Ma:2013yga} we generalize that model by allowing $Z'$ family-dependent charges. 
 In table~\ref{tab:pcontent} $l_{L_i}$,
$e_{R_i}$,$q_{L_i}$,$u_{R_i}$
and $d_{R_i}$ represent the SM fields associated with the $i$-th family.
In order to have non-trivial solutions our model contains, two scalar doublets $H_{1,2}$ (as in the 
original model) and  three  scalar doublets  $\Phi_i$ one for each family~(two additional
fields when comparing with the original reference).
We also have three heavy fermion fields $N_{i}$, which have vector couplings under the extra $U'(1)$  
 abelian gauge symmetry and  singlets under the SM gauge group.

\begin{table}[!h]
\centering
\bgroup                    % nuevo    
\def\arraystretch{1.2}% nuevo
\begin{tabular}{c c c c r c }
\hline\hline \quad Field \quad &\quad Spin\quad &\quad $SU(3)_{c}$
\quad
 &\quad  $SU(2)_{L}$ \quad  & $U(1)_{Y}$& \quad  $U(1)^{\prime}$ \quad  \\
\hline
$l_{Lr}$ & $\frac{1}{2}$ & 1& 2& $-\frac{1}{2}$& $\lr$ \\
$e_{Rr}$ & $\frac{1}{2}$ & 1& 1& $-1$          & $\er $ \\
$q_{Rr}$ & $\frac{1}{2}$ & 3& 2& $\frac{1}{6}$ & $\qr $ \\
$u_{Rr}$ & $\frac{1}{2}$ & 3& 1& $\frac{2}{3}$ & $\ur $ \\
$d_{Rr}$ & $\frac{1}{2}$ & 3& 1& $-\frac{1}{3}$& $\dr $ \\
$N_{(L,R)_r}$ & $\frac{1}{2}$& 1& 1& 0           & $\nnr $ \\
$\Phi_r$ & 0& 1& 2& $\frac12$                  & $\hr $ \\
$H_{1,2}$    & 0& 1& 2& $\frac12$           & $\eta_{1,2}$ \\
\hline\hline
\end{tabular}
\egroup
\caption{Particle content and quantum numbers under the  $SU(3)_C \otimes SU(2)_L\otimes U(1)\otimes U(1)'$  gauge group. 
The indices $r=1,2,3$ run over the three families}
\label{tab:pcontent}
\end{table}

For the $SU(2)_L\otimes U(1)\otimes U(1)'$ symmetry the non-trivial anomaly equations are:

\begin{align}\label{eq:anomalies}
%[SU(2)]^2U(1)' :\hspace{0.2cm}&\frac{1}{3}\Sigma l + \Sigma q=0,\notag\\
[SU(2)]^2U(1)' :\hspace{0.2cm}&0=\Sigma q+\frac{1}{3}\Sigma l,\notag\\
%[SU(3)]^2U(1)':\hspace{0.2cm}&2\Sigma q-\Sigma d-\Sigma u=0,\notag\\
[SU(3)]^2U(1)':\hspace{0.2cm}&0=2\Sigma q-\Sigma u-\Sigma d,\notag\\
%[\text{grav}]^2U(1)':\hspace{0.2cm}&3\Sigma d+3\Sigma u-6\Sigma q+\Sigma e+\Sigma \nu-2\Sigma l=0\notag\\
[\text{grav}]^2U(1)':\hspace{0.2cm}&0=6\Sigma q-3(\Sigma u+\Sigma d)+2\Sigma l-\sigman-\Sigma e +\Sigma N\notag\\
% [U(1)]^2U(1)':\hspace{0.2cm}& \Sigma l-2\Sigma e
% -\frac{2}{3}\Sigma d 
% -\frac{8}{3}\Sigma u
% +\frac{1}{3}\Sigma q\notag\\
[U(1)]^2U(1)':\hspace{0.2cm}& 0=
 \frac{1}{3}\Sigma q
-\frac{8}{3}\Sigma u
-\frac{2}{3}\Sigma d 
+\Sigma l-2\Sigma e
\notag\\
% U(1)[U(1)']^2:\hspace{0.2cm}&
% 2\Sigma u^2-\Sigma d^2-\Sigma q^2+\Sigma l^2-\Sigma e^2=0,\notag\\
U(1)[U(1)']^2:\hspace{0.2cm}&
0=\Sigma q^2
-2\Sigma u^2+\Sigma d^2
-\Sigma l^2+\Sigma e^2,\notag\\
[U(1)']^3:\hspace{0.2cm} & 
0=6\Sigma q^3-3(\Sigma u^3+\Sigma d^3)+2\Sigma l^3-\sigman^3-\Sigma e^3+\Sigma N^3\notag\ .\\
\end{align} 
where $\Sigma f=f_1+f_2+f_3$.
We have also considered the constraints coming from Yukawua interaction terms
\begin{align*}
\centering
\mathcal{L}_Y \supset &\sum_{r=1,2,3} \bar{l}_r\Phi_r e_{r}+\bar{q}_r\tilde{\Phi}_r u_r+\bar{q_r}\Phi_r d_r
+\mu_{N_r}^2\bar{N}_{R_r}N_{L_r}
+\lambda(\Phi_1^\dagger H_1)(\Phi_1^\dagger H_2)+\text{H.C}
\end{align*}
In addition to this Lagrangian, 
by following~\cite{Ma:2013yga}, we also added
the restrictions that come from the terms
$\sum_{r=j,k}\nu_{Lr}\eta_1^{0}\bar{N}_{R_r}$, 
            $\sum_{r=j,k}N_{rL}\nu_{Lr}\eta_2^{0}$,  
where, $H^T_{1,2}=(\eta^+_{1,2},\eta^0_{1,2})$.
\begin{figure}[!h]
    \centering
    \includegraphics[width=5.5cm]{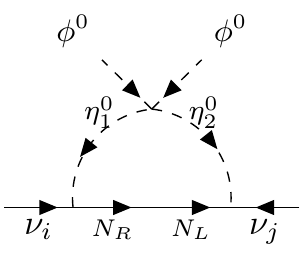}
    \caption{
    One-loop neutrino self-energy.
    }
    \label{fig:one-loop}
\end{figure}
From these terms we obtain  the constraints  
 \begin{align}\label{eq:yukawa}
\quad e_r - l_r +\phi_r  &= 0, \hspace{0.5cm} 
      d_r - q_r +\phi_r  = 0,   \hspace{0.5cm}
      u_r - q_r -\phi_r  = 0,   \notag\\ 
 \lj(\lk)+\eta_1-\nj(\nk)&=0,\hspace{0.5cm}   
  \lj(\lk)+\eta_2+\nj(\nk)=0,\hspace{0.5cm}+\eta_1 +\eta_2-2\phi_{j,k}=0  
   \end{align}
There are several options for the choice of the indices;
in general, the triplet $(ijk)$ is a permutation of (123) and $r=1,2,3$. 
It is possible to combine the indices $j,k$ in three different ways, i.e., (1,2), (1,3) or (2,3). 
Each of these options represents a different model.
\section{Results and conclusions}
\label{sec:results}
 The solution of the equations~(\ref{eq:anomalies}) and~(\ref{eq:yukawa}) are shown
 in tables~\ref{tab:1} and \ref{tab:2}.  
The couplings to SM leptons and the scalar fields $\Phi_r$  in table~\ref{tab:1}, 
are the same for the three families; hence, 
for this solution the three fields $ \Phi_r$ can be reduced to only one. 
That is interesting since in this case.
the number of fields is identical to the original model~\cite{Ma:2013yga} even though the model 
is not universal.
For $ q_1 = q_2 = q_3 = 0 $ and $\etai = 1$
from this solution it is possible to obtain
the original model~\cite{Ma:2013yga}. 
 \begin{table}
 \begin{center}
 \bgroup                    % nuevo    
 \def\arraystretch{1.3}% nuevo
 \begin{tabular}{|c|l|l|l|}
 \hline  
 $f$       &\multicolumn{3}{|c|}{ $\epsilon^{Z'}(f)$}   \\            
 \hline
 \hline
 $l_r$     &$-\sigmaq$         &$\phi_r$   & $+\sigmaq$ \\ 
 $e_r$     &$-2\sigmaq$        &$n^j$      &$+\eta_1-\sigmaq$\\
 $u_r$     &$+q_r+\sigmaq$     &$n^k$      &$+\eta_1-\sigmaq$\\
 $d_r$     &$+q_r-\sigmaq$     &$\eta_2$   &$-\eta_1+2\sigmaq  $\\
 \hline
 \end{tabular}
 \egroup
 \caption{Universal solution in the leptonic sector for the  $Z'$ charges in the equations~(\ref{eq:anomalies}) and~(\ref{eq:yukawa}).
 The free parameters are: the left-handed couplings to the quarks $q_k$,  the heavy-vector field $n_i$,
 and the coupling to the $H_1$ field $\etai$. 
  The integers  $(ijk)$ are a permutation of $(123)$ and they correspond to the family number.} 
 \label{tab:1}
 \end{center}
 \end{table}
The solution in the table~\ref{tab:2} is more interesting 
for flavor physics, due to the fact that there is less dependence 
between the $Z'$ charges of the particles in different families.
  \begin{table}
 \begin{center}
 \bgroup                    % nuevo    
 \def\arraystretch{1.3}% nuevo
 \begin{tabular}{|c|l|l|l|l|l|}
 \hline  
 $f$       &\multicolumn{5}{|c|}{ $\epsilon^{Z'}(f)$}   \\            
 \hline
% \hline
$l_i$     &$-3\qi$    &$l_j$    &$-\frac{3}{2}\left(\qj+\qk\right)  $ &$l_k$      &$-\frac{3}{2}\left(\qj+\qk\right)$ \\
$e_i$     &$-6\qi$    &$e_j$    &$          -3\left(\qj+\qk\right)  $&$e_k$      &$-3\left(\qj+\qk\right)$\\
$u_i$     &$+4\qi$    &$u_j$    &$+\frac{1}{2}\left(5\qj+3\qk\right)$&$u_k$      &$+\frac{1}{2}\left(3\qj+5\qk\right)$\\
$d_i$     &$-2\qi$    &$d_j$    &$-\frac{1}{2}\left(\qj+3\qk\right) $&$d_k$      &$-\frac{1}{2}\left(3\qj+\qk\right) $\\
\hline
$\hi$     & $+3\qi$   &$\hj$    &$+\frac{3}{2}\left(\qj+\qk\right)  $&$\hj$      &$+\frac{3}{2}\left(\qj+\qk\right)$\\
% $N_R^i$   &$+N_L^i$   &$N_R^j$  &$+N_L^j$&$N_R^j$    &$+N_L^j$\\
$\etaj$   &$-\etai+3(\qj+\qk)$  &$\nj$  &$+\frac{1}{2}\left(2\etai-3\qj-3\qk\right)$  &$\nk$    &$+\frac{3}{2}\left(2\etai-3\qj-3\qk\right)$\\  
 \hline
 \end{tabular}
 \egroup
 \caption{Solution for the  $Z'$ charges in the equations~(\ref{eq:anomalies}) and~(\ref{eq:yukawa}).
 The free parameters are: the  left-handed couplings to the quarks $q_k$,  the heavy-vector field $n_i$,
 and the coupling to the $H_1$ field $\etai$. 
  The integers  $(ijk)$ are a permutation of $(123)$ and they correspond to the family number.} 
 \label{tab:2}
 \end{center}
 \end{table}
By choosing $(i,j,k)=(1,2,3)$ and   $q_1=q_2$, it is possible to obtain a 
model with generation dependent charges in the lepton sector. 
By a convenient mixing matrix for the right-handed fields, 
it is possible to obtain a model without flavor changing neutral currents. 
This model is also important since it represents a realization of 
a non-universal $Z'$ model without  right-handed neutrinos with 
 allowed one-loop contributions to the neutrino masses. 
As explained in reference~\cite{Ma:2013yga} the dark matter candidate is either,
the real part of $\eta^0_{1,2}$ or one of the exotic fermions $N_i$. 
 
\FloatBarrier
\section*{Acknowledgment}
We thank Financial support from \enquote{Patrimonio Autónomo 
Fondo Nacional de Financiamiento para la Ciencia, la Tecnología
 y la Innovación, Francisco José de Caldas}, 
 and \enquote{Sostenibilidad-UDEA}.

% \bibliographystyle{apsrev4-1longdoi}
% %\bibliographystyle{apsrev4-1long}
% \bibliography{biblio}

\end{document}